\documentstyle[prl,aps,epsbox]{revtex}
\begin{document}
\draft
%\preprint{\vbox{\hfill {\bf 1st DRAFT} \today}}
\preprint{}
\twocolumn[\hsize\textwidth\columnwidth\hsize\csname@twocolumnfalse%
\endcsname
\title{
Wave Function Shredding
by Sparse Quantum Barriers}

\author{
Taksu Cheon${ }^{1}$, 
Pavel Exner${ }^{2,3}$, and  
Petr {\v S}eba${ }^{1,3,4}$
}
\address{
 ${ }^1$ Laboratory of Physics, Kochi University of Technology,
 Tosa Yamada, Kochi~782-8502, Japan \\
% ${ }^2$ Theory Group, High Energy
% Accelarator Research Organization (KEK), Tanashi, Tokyo~188-8501,
% Japan \\
 ${ }^2$ Nuclear Physics Institute, Academy of Sciences of the Czech
 Republic, 25068~\v{R}e\v{z} u Prahy, Czech Republic \\
 ${ }^3$ Doppler Institute, Czech Technical University,
 B\v{r}ehov\'{a} 7, 11519~Praha, Czech Republic \\
 ${ }^4$ Institute of Physics, Academy of Sciences of the Czech
 Republic, Cukrovarnick\'{a} 10, 16253~Praha, Czech Republic }
%
%
%\date{\today}
\date{April 14, 2000}
\maketitle
%
%\widetext
%
\begin{abstract}
%------------------%
We discuss a model in which a quantum particle passes through 
$\delta$ potentials arranged in an increasingly sparse way. For
infinitely many barriers we derive conditions, expressed in terms
ergodic properties of wave function phases, which ensure that the
point and absolutely continuous parts are absent leaving a purely
singularly continuous spectrum. For a finite number of
barriers, the transmission coefficient shows extreme sensitivity
to the particle momentum with fluctuation in many different scales. 
We discuss
a potential application of this behavior for erasing the information 
carried by the wave function.
%------------------%
%
%
%\vspace*{2mm}
%\noindent KEYWORDS:
%  one-dimensional system,
%  quantum transport,
%  sparse potential,
%  singular spectrum,
%  sensitive dependence on initial state
%------------------%
\end{abstract}
\pacs{PACS Nos: 3.65.-w, 2.30.Sa, 5.45.-a, 5.60.Gg}
%
% %
]
%
% %
%--------------------------------------------------------
%\twocolumn 
\narrowtext
%
%

%------------------%
The prospective use of quantum states as information carriers
%\cite{SH97} 
put new emphasis on tailoring systems which could
process the wave function in a desired way. One of such devices,
potentially of crucial importance, is an information eraser, or
briefly {\em shredder.} If we look for ways in which a quantum
system could achieve this goal, a natural idea is to seek
situations where a slight change of the input is magnified
significantly in the output.

In classical mechanics, the sensitive dependence on the initial
condition of that kind is the essence of the deterministic
randomness exhibited by non-integrable dynamical systems
\cite{OT93}. In quantum mechanics, on the other hand, such
sensitive dependence has been an elusive object despite many
efforts, and a widely accepted counterpart to the mentioned
classical behavior is missing. 
Instead, ``quantum signatures'' of
chaos are often studied which typically manifest themselves
through statistical properties of spectra of various observables
\cite{RE92}.
There has been several attempts to find sensitive dependence in
quantum mechanics using time periodic models \cite{GK91,SA98}.
However, they are not easy to solve and the numerical
results often contain ambiguities. 
In this letter, 
we abandon such traditional schemes, and instead 
propose a simple
one-dimensional model with a potential which has many different
length scales.

We start with an infinite number of barriers whose distances
grow,
and ask for conditions under which its spectrum
exhibits irregularity, defined in mathematical terms as being
singularly continuous.
We will have unexpected encounter with an associated classical
system whose ergodicity brings about the appearence of
purely singularly continuous spectrum. 
We then consider a scattering system in which the
number of barriers is finite. Through numerical analysis, it is
shown that the transmission
coefficient exhibits a highly irregular behavior. 
We also show that a wave packet passing through the system
undergoes randomizing during which the shape
information contained in the input wave function gets lost.

%------------------%

We choose an explicitly solvable model of a particle on the line
interacting with $(2N+1)$ $\delta$ potentials having the same
coupling constant, which is described by the stationary
Schr\"odinger equation
%------------------%
\begin{eqnarray}
\label{schrod}
 \left[ -{{d^2} \over {dx^2}}
 +\sum_{n=-N}^{N}{v \delta(x-x_n)}
 \right]
 \varphi(x) = k^2 \varphi(x) ,
\end{eqnarray}
%------------------%
where the positions $x_i$ are indexed in
ascending order.
% $x_{n}$ $< x_{n+1}$ $(n = -N .. {N-1})$.
Properties of such systems depend crucially 
on the positioning of
the $\delta$ interactions. 
In our case they are arranged
symmetrically, $x_{-n}=-x_n$ with $x_0=0$, and in such a way that
their positions are increasingly sparse;
%------------------%
\begin{eqnarray}
\label{spars}
 x_{|n|+2} - x_{|n|+1} > 
 x_{|n|+1} - x_{|n|} \,.
\end{eqnarray}
%------------------%
We are particularly interested in the situations where the
distances grow polynomially, exponentially, or faster,
%------------------%
\begin{eqnarray}
\label{spac1}
% x_{|n|} = b\,|n|^{\beta} + c\,{\mathrm{e}}^{a|n|^\gamma}
 x_{|n|} = b\,|n|^{\beta} + c\,{\rm{e}}^{a|n|^\gamma}
\end{eqnarray}
%------------------%
for $n\ne 0$, with 
suitable positive $a$, $b$, $c$, and $\beta>1$,
$\gamma \ge 1$.

We employ transfer matrices which relate solutions of
Eq.~(\ref{schrod}) at different points. 
%We associate
%conventionally with $\varphi$ the two-component vector
We define the two component vector $\Phi(x)$ 
from the wave function and its derivative as
%------------------%
\begin{eqnarray}
%\label{Phi}
 \Phi(x)= \left( {\matrix{{\varphi  (x)}\cr
                {\varphi '(x)}\cr}}
 \right)\,;
\end{eqnarray}
%------------------%
then the transfer matrix ${\cal M}(x,y)$ is given by
%------------------%
\begin{eqnarray}
%\label{transfer}
 \Phi(x) = {\cal M}(x,y) \Phi(y).
\end{eqnarray}
%------------------%
Because of the zero-range nature of the interaction \cite{AG88},
the transfer matrix ${\cal M}(x,y)$ is expressed as interlaced 
products of ones describing the
free motion with momentum $k$,
%------------------%
\begin{eqnarray}
%\label{free transf}
% {\cal M}^{\mathrm{free}}_k(x,y) = \left( {\matrix{\cos{k(x-y)} &
 {\cal M}^{\rm{free}}_k(x,y) = \left( {\matrix{\cos{k(x-y)} &
 {1\over k}\sin{k(x-y)} \cr
                -k\sin{k(x-y)} & \cos{k(x-y)} \cr}}
 \right),
\end{eqnarray}
%------------------%
and the point interaction transfer matrices
%------------------%
\begin{eqnarray}
%\label{point transf}
 {\cal M}(x_n+0,x_n-0) \equiv {\cal M}_v =
 \left( {\matrix{1 & 0 \cr
                v & 1 \cr}}
 \right).
\end{eqnarray}
%------------------%
We do not index the latter since the $\delta$ potentials are
supposed to be identical. If $x>y$ the component matrices of
${\cal M}(x,y)$ are multiplied in the order reversed to that of
the involved intervals.

To express the amplitudes for the left-to-right scattering over a
finite number of barriers, we suppose
%------------------%
\begin{eqnarray}
\label{l-r}
% \varphi(x) = \left\{ \begin{array}{lcl} {\mathrm{e}}^{ikx}
%   +r(k) {\mathrm{e}}^{-ikx} \; & \dots & \quad  x<x_{-N} \\
%   t(k) {\mathrm{e}}^{ikx} \; & \dots & \quad  x>x_N
%   \end{array} \right.
 \varphi(x) 
 &=& {\rm{e}}^{ikx}
   +r(k) {\rm{e}}^{-ikx}   
      \quad  x<x_{-N} 
\\ \nonumber
 &=&  \quad \quad \quad
    t(k) {\rm{e}}^{ikx}  
      \quad \ \   x>x_N
\end{eqnarray}
%------------------%
and match the boundary values at $x_{\pm N}$ using the matrix
${\cal M}\equiv {\cal M}(x_N+0,x_{-N}-0)$; then an easy
computation using $\det{{\cal M}}=1$ gives, in particular,
%------------------%
%\begin{eqnarray} %\label{r}
% r(k) &\!=\!&  -\,\frac{{\cal M}_{21} + ik({\cal M}_{22}
%   \!-\!{\cal M}_{11}) + k^2{\cal M}_{12}}
%   {{\cal M}_{21} - ik({\cal M}_{22} \!+\! {\cal M}_{11})
%   - k^2{\cal M}_{12}}\,, \\
\begin{equation}
\label{t}
 t(k)
% &\!=\!& -\,
   = \frac{2ik}{{\cal M}_{21}
   - ik({\cal M}_{22} \!+\! {\cal M}_{11})
   - k^2{\cal M}_{12}}\,. 
\end{equation}
%\end{eqnarray}

%------------------%
%
Before making use of the last formula, let us look what happens in
the limit $N\to\infty$ when the scattering loses meaning; 
we focus on the change of the
spectral characteristics of the system.
Traditionally, most attention has been paid in 
physics literature to two types of spectra
the point and the absolutely continuous. 
%({\it ac}).
Recently, there has been a growing recognition of 
the importance of the third type,
which is mathematically well studied,
the singularly continuous 
%({\it sc}) 
spectrum. 
Here, we ask ourselves under which
condition the spectrum of our system  
is purely singularly continuous. 
In view of the symmetry it is sufficient to consider
the operator
%------------------%
\begin{eqnarray}
\label{H_theta}
 H_{\vartheta} = -{{d^2} \over {dx^2}}
 +\sum_{n=1}^{\infty}{v \delta(x-x_n)}
\end{eqnarray}
%------------------%
on $L^2(0,\infty)$ with the boundary condition
%------------------%
\begin{eqnarray}
%\label{theta bc}
 \varphi(0) \cos\vartheta + \varphi'(0) \sin\vartheta = 0\,,
\end{eqnarray}
%------------------%
in particular, our case with the $\delta$ potential of strength
$v$ at the origin corresponds to $\vartheta= -\arcsin(1+
{v^2\over 4})^{-1/2}$.

Several recent papers investigated the relations between the
spectrum of the operator (\ref{H_theta}) and growth properties of
the corresponding transfer matrices. A sufficient condition for
the absence of the point spectrum is \cite{SS96}
%------------------%
\begin{equation} \label{sis cond}
 \int_0^{\infty} {dx\over \|{\cal M}(x,0)\|^2} = \infty\,.
\end{equation}
%------------------%
The norm of the transfer matrix can be estimated by the product of
norms of its component matrices,
%------------------%
\begin{eqnarray} \label{norm est}
{\|{\cal M}(x_n,0)\|}
 \le \nu^2 
%\left( {|v|\over 2} + \sqrt{{v^2\over 4} +1} \right)^n
 \left( \max(k,k^{-1}) \right)^n \equiv {\cal M}[v,k]^n,
\end{eqnarray}
%------------------%
with
%------------------%
\begin{eqnarray} \label{dfnu}
\nu :={|v|\over 2} + \sqrt{{v^2\over 4} +1} 
%\nu :=(|v|/ 2) + \sqrt{(v^2/ 4) +1} 
\end{eqnarray}
%------------------%
Denoting $\Delta_n:= x_n-x_{n-1}$, we have
%------------------%
\begin{equation} \label{sis est}
 \int_0^{x_N} {dx\over \|{\cal M}(x,0)\|^2} \ge \sum_{n=1}^N
 {\Delta_n \over \|{\cal M}[v,k]\|^{2n}}\,,
\end{equation}
%------------------%
so the integral diverges, for example, if one has
$\Delta_n \ge c_1 n^{-1/2}\,{\rm{e}}^{a n}$
%------------------%
%\begin{eqnarray}
%\label{spac2}
% \Delta_n \ge c_1 n^{-1/2}\,{\mathrm{e}}^{a|n|}
% \Delta_n \ge c_1 n^{-1/2}\,{\rm{e}}^{a|n|}
%\end{eqnarray}
%------------------%
for some $c_1>0$ and
%------------------%
\begin{eqnarray}
%\label{increment}
 a > 2\left\{ \ln\, \max(k,k^{-1}) 
%+ \ln \left( {|v|\over 2}+ \sqrt{{v^2\over 4} +1} \right) 
+ \ln \nu
\right\}.
\end{eqnarray}
%------------------%
If we want a condition independent of the parameters we have to
require a slightly faster growth, for instance, that of
(\ref{spac1}) with any $\gamma>1$.

Sparse
potentials with growing $\Delta_n$ are
known to be good candidates to produce singularly continuous
spectra. Already the first example by Pearson \cite{PE78} is of
this kind, many others can be found in \cite{SS96,KLS97}.
In distinction to a typical Pearson potential, however, the
height of the barriers does not decrease here. This makes it
more difficult to exclude the absolutely continuous spectrum. To
show where is the core of the problem we employ the modified
Pr\"ufer 
%(or EFGP) 
technique \cite{KLS97}.
% which can be traced back
%to \cite{E72}, \cite{GP75}. 
It is based on the Ansatz
%------------------%
\begin{equation} %\label{Ansatz}
 \varphi(x) = R(x) \sin\theta(x)\,,\quad \varphi'(x) = k R(x)
 \cos\theta(x)\,.
\end{equation}
%------------------%
The amplitude $R(x)$ is constant
in the absence of a potential, so the solution
of the Schr\"odinger equation in the interval $(x_{n-1},x_n)$
between the barriers  is
%------------------%
\begin{equation} %\label{EFGP soln}
 \varphi(x) = R_n \sin\left[\theta_n +k(x-x_{n-1})\right] \,.
\end{equation}
%------------------%
The $\delta$ coupling at the point $x_n$ means the wave function
continuity together with
%------------------%
\begin{equation} %\label{delta bc}
 \varphi'(x_n+0) - \varphi'(x_n-0) = v \varphi(x_n)\,.
\end{equation}
%------------------%
This yields a system of equations for $R_n, R_{n+1}$ which is
solvable under the Eggarter condition\cite{E72,GP75}
%------------------%
\begin{equation} \label{Egga}
 \cot\theta_{n+1} = \cot\left[\theta_n + k\Delta_n\right] +
 {v\over k}\,.
\end{equation}
%------------------%
This determines the phase of the wave function in an iterative
way.
%however, we will not need the values of $\theta_n$ in the
%argument below. 
%Solving now the equations for $R_n$ we get
For the ratio of the amplitudes  $r_n$ $:= R_{n+1}/R_n$,
we get
%------------------%
\begin{eqnarray} 
%\label{ratio1}
r_n^2
= {\sin^2\left[\theta_n + k\Delta_n\right] \over
 \sin^2\theta_{n+1}}
= 
 \mu + \sqrt{\mu^2-1}\sin{\beta_n}
\end{eqnarray}
%------------------%
with the definitions
%------------------%
\begin{eqnarray} 
\label{dfbet}
 \beta_n := 2\theta_n + 2 k \Delta_n- \alpha,
\end{eqnarray}
%------------------%
and
%------------------%
\begin{eqnarray} 
\label{dfalmu}
\mu := 1+{{v^2}\over{2k^2}},
\quad
\alpha := \arccos\left( 1+{v^2\over4k^2}\right)^{-1/2} .
\end{eqnarray}
%------------------%
The new phase variable $\beta_n$ is determined by
the evolution
\begin{equation} 
\label{Eggamd}
 \cot \left(
{{\beta_{n+1}+\alpha}\over2}-k\Delta_{n+1} 
      \right)
= \cot\left({{\beta_n +\alpha}\over 2}\right)
 +
 {v\over k} .
\end{equation}
%------------------%
By Theorem~1.1 of \cite{KLS97}, the absolutely continuous spectrum
will be absent if $\|{\cal M}(x,0)\|$ tends to infinity at least
for some sequence of points and {\it a.e.} $k$. Since
%------------------%
\begin{equation} 
%\label{norm est}
 \|\Phi(x)\|^2 \ge \min(1,k) R(x)^2,
\end{equation}
%------------------%
it is enough to demonstrate the growth for a subsequence of
$\{R_n\}$, or equivalently, that the quantity
%------------------%
\begin{equation} 
%\label{ratio3}
 \sum_{n=1}^N \ln r_n^2
\end{equation}
%------------------%
is exploding for a subsequence of the indices $N$.
% Here it is
%useful to take ergodic properties of the oscillating quantity at
%the r.h.s. of (\ref{ratio2}) into account. Recall that by
Now, for a continuous function $F\in C([-\pi,\pi])$, one has
%------------------%
\begin{equation} %\label{ergodic}
\lim_{n\to\infty} {1\over n}\, \sum_{k=1}^n F(\beta_n) =
\int_{-\pi}^{\pi} F(\beta)\, d\beta
\end{equation}
%------------------%
if the sequence $\{\beta_n\}_{n=1}^{\infty}$ is
{\em uniformly distributed} in $[-\pi,\pi]$ \cite{cfs}.
In our case, we have $F(\beta_n):= \ln r_n^2$.
Using the fact that the sine is an odd function we find
%------------------%
\begin{equation} %\label{posit}
\int_{-\pi}^{\pi} F(\beta)\,d\beta 
= \int_0^{\pi} \ln \left( \mu^2 \cos^2\beta
+ \sin^2 \beta \right)\, d\beta > 0
\end{equation}
%------------------%
for any $\mu>1$. It follows that the sequence $\{R_N\}$ is
exponentially growing if $\{\beta_n\}$ is uniformly
distributed.
Thus the problem of singularly continuous 
quantum spectrum is turned into that of
dynamical properties of the {\em classical map} (\ref{Eggamd}).
%%%%%%%%%%%%%%%
\begin{figure}
\center
\psbox[hscale=0.55,vscale=0.55]{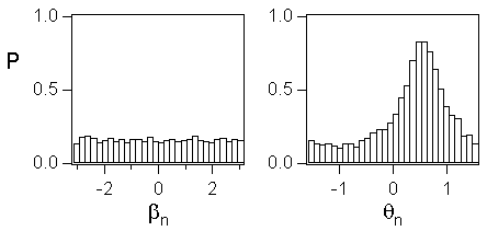}
\vspace{3mm}
\caption{(a) The histogram shows the distribution of the 
phase
$\beta_n$ in the interval $[-\pi,\pi)$ for $v=k=1$ after
$N=20000$ iterations. (b) The distribution of $\theta_n$ for the
same values.
}
\end{figure}
%%%%%%%%%%%%%%%
We shall not attempt to prove this property and limit ourselves to
illustrating it for $x_n = n^2$. The results are shown in Fig.~1;
similar behavior may be observed for other values $v$, almost all
\cite{almost} values of the momentum $k$, as well as for
$x_n = n^{\alpha}$ with $\alpha > 1$. 
Although the original phases
$\theta_n$ may not be uniformly distributed as
shown in the Fig. ~1, it is sufficient that
$\{\theta_n\}$ is ergodic, if
the sequence $\{2k\Delta_n\}$ is
uniformly distributed, which turns out
to be the case when  $k$ is an irrational 
multiple of $\pi$, for example, 
for $x_n =n^2$ \cite{cfs}.
%, Sec. 7.2.

Let us sort out the above discussion. If the number of barriers
is infinite the absolute continuity for Hamiltonians of the type
(\ref{schrod}) requires a particular arrangement, e.g., a periodic
one. On the other hand, randomly distributed $\delta$ barriers
yield almost surely a pure point spectrum by Anderson localization
\cite{PF92}. The exclusive character of absolutely continuous 
spectra makes
easier to get rid of them; the above results suggests that a
powerlike sparseness may be sufficient. At the same time, the
borderline between dense point and singularly continuous spectra
is extremely unstable and an exponential or faster growth may be
needed to get a purely singularly continuous spectrum.

%------------------%
%
Let us return to systems with a finite $N$.
When the number of barrriers is finite, 
the positive energy spectra of
the system is purely absolutely continuous.
However, one expects certain trace of
singular behavior in observable quantities
even with finite $N$.  One suspects, for example, 
that the poles of the scattering matrix is distributed 
differently for sparse potentials, which should
be reflected in the increasingly wild behavior of scattering 
amplitudes as one increase the sparseness of the potential
with fixed finite $N$.
%%%%%%%%%%%%%%%
\begin{figure}
\center
\psbox[hscale=0.55,vscale=0.55]{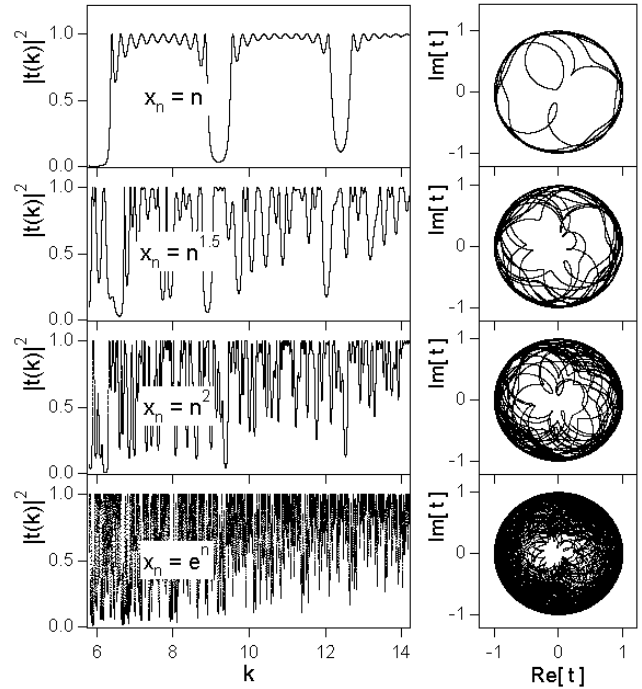}
\vspace{3mm}
\caption{Transmission probability and transmission amplitude over
$2N+1$$=11$ barriers for different sparseness.
}
\end{figure}
%%%%%%%%%%%%%%%
That expectation is confirmed in Figure 2, in which we
plot the transmission coefficient $|t(k)^2|$ as a function
of incident momentum $k$.  The number of barriers is
chosen to be  $2N+1 = 11$. One has increasing sparseness
from top to bottom. 
Irregular fluctuation of the transmission coefficient
which is reminiscent of the {\it classical} chaotic
scattering \cite{OT93}
is clearly observed for sparse potentials.
Note also the ``ergodic'' behavior of the trajectory
of $\{ {\rm Re} t(k), {\rm Im} t(k) \}$ with increasing sparseness.
In Figure 3, we plot the transmission coefficient
for a larger number of barriers $2N+1=21$ 
with exponential sparseness $x_n=e^n$.
The top graph is magnified by factor of 10
in the second graph, part of which is magnified again by
factor 10 in the third graph. 
From this, one can discern
the fact that the irregular fluctuation of quantum
scattering matrix takes place in many diffrent scale.
Notice that the despite the irregularity the systems
of sparse barriers discussed here still exhibit a considerable
transmission: for the same number of {\em randomly positioned}
barriers Anderson localization takes place and $|t(k)|^2$ is
practically identical to zero.
%------------------%
%
%%%%%%%%%%%%%%%
\begin{figure}
\center
\psbox[hscale=0.55,vscale=0.55]{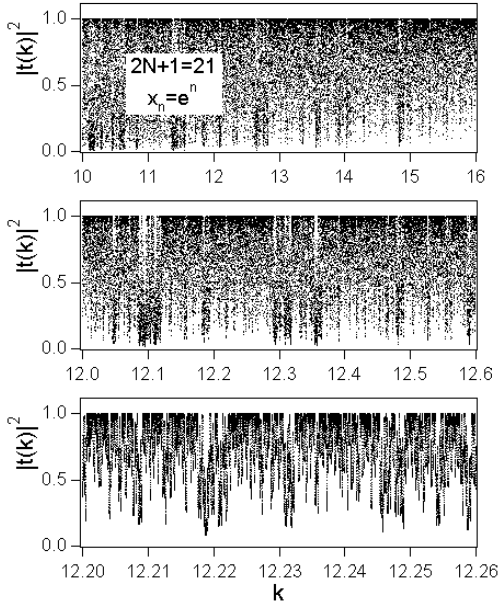}
\vspace{3mm}
\caption{Transmission amplitude $|t(k)|^2$ for exponentially
sparse 21 barriers in diffrent $k$ scale.
}
\end{figure}
%%%%%%%%%%%%%%%

In actual experimental settings, it is often 
difficult to prepare the particle beam 
of single wavelength.
One would rather consider the problem as
the transmission of {\it wave packet}.
When one has the incoming wave with 
Fourier component
$\varphi(k)$, the outcoming wave will have 
the spectra $t(k) \varphi(k)$.
%Suppose one puts in 
%the wave packet $\varphi_{in}(x,t)$
%given by
%------------------%
%\begin{eqnarray}
%\label{14} 
%\varphi_{in}(x,t) = {1\over{2\pi}} \int_{-\infty}^{\infty}
%{e^{-ikx-ik^2t}\varphi_{in}(k)\, dk}\,,
%\end{eqnarray}
%------------------%
The initial wave packet $\varphi_{in}(x,t)$
will be proccessed after the transmission to give
the outcomming wave packet
%------------------%
\begin{eqnarray}
\label{convol} 
\varphi_{out}(x,t) = \int_{-\infty}^{\infty}
{t(x-y)\varphi_{in}(y,t)\, dy} ,
\end{eqnarray}
%------------------%
where $t(x)$ is the Fourier transform of 
%the transmission amplitude 
$t(k)$.
%, i.e.
%%------------------%
%\begin{eqnarray}
%\label{16} t(x) = {1\over{2\pi}} \int_{-\infty}^{\infty} {e^{-ikx}
%t(k)\, dk} .
%\end{eqnarray}
%%------------------%
Because of the fluctuation in all scale
is present in $t(k)$, the quantity $t(x)$ 
%is again a highly irregular function.
%However, because of the folding, this irregular
%fluctuation is smeared out in the 
%outcomg wave packet $\varphi_{out}(x,t)$ when the
%spectrum $\varphi_{in}(k)$ is not sufficiently
%monocromatic.
%What is even more important, the Fourier transform $t(x)$ 
has a slow decay at large distances.  
As a consequence, any information contained in
the shape of the wave packet
$\varphi_{in}(x,t)$ 
is effectively washed out in $\varphi_{out}(x,t)$
after the convolution (\ref{convol}). 
%------------------%
%

In summary, we have analyzed transport of a quantum particle
through a sparse family of $\delta$ barriers of identical
strength. If the barriers are arranged in a 
non-random way there is
a non-negligible transmission. 
When the barriers are placed sparsely, the transmission
coefficient shows an irregular dependence on incoming
momentum in self-similar manner, 
and the shape of wave packets passing through the system is
randomized.
%Its randomizing
%effect on the wave packet is pointed out.
Properties of the
transport can be studied through spectral properties of infinite
barrier systems. 
Because of the solvability of the problem, a rather detailed 
analysis can be carried out, and a hidden relation
between the spectral properties of the system 
and the ergidicity of a certain classical dynamical system
is revealed. 
We stress that the simple model considered here
is not so far removed from experimentally realizable system with
existing technology.
Thus we conclude that the realization of 
%truly chaotic, 
initial-condition sensitive,
information erasing process in purely quantal setting may
be already within our reach.

\bigskip
%\newpage
%\acknowledgements
%\vspace*{5mm}
%
TC expresses
his thanks to Prof. 
Toshiya Kawai and Prof. Izumi Tsutsui 
for helpful dicsussions.
This work has been
supported in part by the Grant-in-Aid (No. 10640396) of the
Japanese Ministry of Education and Grant ME170 of the Czech
Ministry of Education.
%

%\bigskip
%\newpage
%
%------------------%

%------------------%
%

\begin{references}
\bibitem{almost}
There are values of $k$ where this property is violated such as
entire fractions   of $\pi$.
\end{references}
\end{document}